\documentclass[aps, prd, twocolumn, preprintnumbers, nofootinbib, superscriptaddress, balancelastpage, 10pt]{revtex4-1}

\usepackage[utf8]{inputenc}
\usepackage{graphicx}
\usepackage{amsmath, amsfonts, amssymb}
\usepackage{bm,bbm}
\usepackage{enumitem}
\usepackage{hyperref}
\usepackage{xcolor}

\newcommand{\beq}{\begin{equation}\begin{aligned}{}}
\newcommand{\eeq}{\end{aligned}\end{equation}}
\newcommand{\beqa}[1]{\begin{equation}\begin{aligned}{#1}}
\newcommand{\eeqa}{\end{aligned}\end{equation}}

\newcommand{\del}{\partial}

\newcommand{\smallsubsubsection}[1]{\vspace{-1.5em}\subsubsection{#1}\vspace{-1em}}

\begin{document}

\title{Heavy QCD Axion at Belle II: Displaced and Prompt Signals}

\preprint{KEK-TH-2343}

\author{Emilie Bertholet}
\affiliation{Tel Aviv University, School of Physics and Astronomy, Tel Aviv, 69978, Israel}

\author{Sabyasachi Chakraborty}
\affiliation{Department of Physics, Florida State University, Tallahassee, FL 32306, USA}
\affiliation{SISSA International School for Advanced Studies, Via Bonomea 265, 34136, Trieste, Italy}

\author{Vazha Loladze}
\affiliation{Department of Physics, Florida State University, Tallahassee, FL 32306, USA}

\author{Takemichi Okui}
\affiliation{Department of Physics, Florida State University, Tallahassee, FL 32306, USA}
\affiliation{Theory Center, High Energy Accelerator Research Organization (KEK), Tsukuba 305-0801, Japan}

\author{Abner Soffer}
\affiliation{Tel Aviv University, School of Physics and Astronomy, Tel Aviv, 69978, Israel}

\author{Kohsaku Tobioka}
\affiliation{Department of Physics, Florida State University, Tallahassee, FL 32306, USA}
\affiliation{Theory Center, High Energy Accelerator Research Organization (KEK), Tsukuba 305-0801, Japan}

\begin{abstract}
The QCD axion is a well-motivated addition to the standard model to solve the strong $C\!P$ problem. If the axion acquires mass dominantly from a hidden sector, it can be as heavy as $O(1)$~GeV, and the decay constant can be as low as $O(100)$~GeV without running into the axion quality problem. We propose new search strategies for such heavy QCD axions at the Belle~II experiment, where the axions are expected to be produced via $B\to K a$. We find that a subsequent decay  $a\to 3\pi$ with a displaced vertex leads to a unique signal with essentially no background, and that a dedicated search can explore the range $O(1$--$10)$~TeV of decay-constant values. We also show that $a\to \gamma\gamma$ can cover a significant portion of currently unexplored region of $150 \lesssim m_a \lesssim 500$~MeV.    
\end{abstract}

\maketitle
\section{Introduction}
The axion is one of the most well-motivated hypothetical particles beyond the standard model (SM) of particle physics. The original axion was predicted by Weinberg and Wilczek~\cite{Weinberg:1977ma, Wilczek:1977pj} as the pseudo-Nambu-Goldstone boson of the spontaneously broken U(1) symmetry proposed by Peccei and Quinn~\cite{Peccei:1977hh, Peccei:1977ur} to solve the strong $C\!P$ problem~\cite{tHooft:1976rip}. While the axion could have additional couplings to the SM particles, the minimal effective Lagrangian thus motivated, up to terms to be included for renormalization, is given by
\begin{align}
\mathcal{L}
=\mathcal{L}_\text{SM}
+\frac{\alpha_s}{8\pi} \frac{a}{f_a} G \widetilde{G}
+\frac{1}{2} (\del_\mu a)^2 - \frac{m_a^2}{2} a^2
\,,\label{eq:treelevellagrangian}
\end{align}
where $a$ is the axion, with mass $m_a$ and decay constant $f_a$, and $G$ is the gluon. 
Such an axion, which we call the \emph{QCD axion}, is the subject of this paper.

Most phenomenological studies and experimental searches for the QCD axion have so far focused on \emph{light} axion masses~\cite{Kim:2008hd,Zyla:2020zbs}, i.e., $m_a < 3m_\pi$, where the physics is dominated by the axion-photon-photon coupling necessarily induced upon QCD confinement, even though the underlying Lagrangian~(\ref{eq:treelevellagrangian}) lacks such coupling.
This coupling provides various experimental handles, such as the $a \to \gamma\gamma$ decay and the axion-photon conversion in a background magnetic field.

There are, however, good reasons to explore \emph{heavy} axion masses, $m_a > 3m_\pi$, where hadronic physics controls the phenomenology. In particular, such a heavy QCD axion can provide a simple solution~\cite{Agrawal:2017ksf} to the \emph{axion quality problem}~\cite{Kamionkowski:1992mf, Holman:1992us,Barr:1992qq, Ghigna:1992iv}, i.e., the violation of the U(1) Peccei-Quinn symmetry by higher dimensional Planck-suppressed operators induced (presumably) by quantum gravity. Such violation should become harmless if $f_a \lesssim O(10)~{\rm TeV}$~\cite{Agrawal:2017ksf}, 
but this would imply $m_a \gtrsim O(1)\>{\rm keV}$ due to the relation $m_a \sim m_\pi f_\pi / f_a$, if the axion mass is  induced solely by QCD\@. This part of the parameter space, however, is excluded by beam dump
experiments~\cite{Bjorken:1988as, Blumlein:1990ay, Bergsma:1985qz} and astrophysical observations~\cite{Anastassopoulos:2017ftl, Raffelt:2006cw, Raffelt:1996wa, Friedland:2012hj}. The simplest way out is to introduce additional contributions to the axion mass, such that $m_a \gg m_\pi f_\pi / f_a$~\cite{Agrawal:2017ksf}.
For our purpose, this simply amounts to treating $m_a$ and $f_a$ in the Lagrangian~(\ref{eq:treelevellagrangian}) as independent parameters. Such treatment can be justified by (small modifications of) many ultraviolet (UV)-complete models, such as those in Refs.~\cite{Fukuda:2015ana, Agrawal:2017eqm, Agrawal:2017ksf, Gaillard:2018xgk, Gherghetta:2020keg}. Also, a heavy QCD axion could be relevant to inflation, and in this case the interesting parameter is $m_a\sim 10^{-6}f_a$ \cite{Takahashi:2021tff}. 

In searching for a heavy QCD axion experimentally, the presence of the $aG\widetilde{G}$ coupling in Eq.~(\ref{eq:treelevellagrangian}) implies that the predominant axion production should be hadronic. As a result, experiments such as proton beam dump~\cite{Bergsma:1985qz, Aloni:2018vki}, kaon decays~\cite{Georgi:1986df,Bardeen:1986yb, Alves:2017avw, Gori:2020xvq, Artamonov:2005ru, Ceccucci:2014oza, Abouzaid:2008xm, Ahn:2018mvc, Bauer:2021wjo}, precision measurements of pion decays~\cite{Aguilar-Arevalo:2019owf,Pocanic:2003pf, Altmannshofer:2019yji}, fixed target~\cite{AlGhoul:2017nbp, Aloni:2018vki, Aloni:2019ruo}, and colliders~\cite{Abbiendi:2002je, Knapen:2016moh, Aloni:2018vki} set strong bounds on $f_a$. For $m_a\gtrsim 50\>{\rm GeV}$ the CMS dijet search excludes large regions of parameter space \cite{Sirunyan:2017nvi, Mariotti:2017vtv}.

On the other hand, the range $O(100)\>{\rm MeV} \lesssim m_a \lesssim 50\>{\rm GeV}$ is poorly constrained. For $m_a>400~{\rm MeV}$ the kaon and beam-dump experiments are not very effective, and in fact, the strongest probe to-date is from $B\rightarrow Ka$. The study of this channel was pioneered in Ref.~\cite{Aloni:2018vki}, where the underlying $b \to sa$ amplitude was estimated and the branching fractions of the subsequent $a$ decay into various final states were inferred by a data-driven method. The leading-order determination of the $b \to sa$ amplitude requires a 2-loop calculation and leading 2-loop renormalization group evolution with varying initial conditions for the evolution. This was performed in Ref.~\cite{Chakraborty:2021wda}, where the result was combined with the data-driven branching fractions of~\cite{Aloni:2018vki} to derive constraints from the past $B$-factory results and future projections for Belle~II for the prompt axion decays $a\to \pi^0\pi^+\pi^-$, $\eta \pi^+\pi^-$, $KK\pi$, and $\phi\phi$. Finally, axion production in $\phi\to\gamma a$ and $\eta' \to \pi\pi a$ also sets constraints on the parameter space~\cite{ParticleDataGroup:2020ssz}.

In this paper, we aim to significantly extend the study of prospects of future searches at Belle~II, which was   discussed in Ref.~\cite{Chakraborty:2021wda}. 
In particular, we show that displaced axion decays to $\pi^0\pi^+\pi^-$ should provide a powerful search strategy, since the axion tends to be long-lived in the parameter space of our interest, and hence this signature is associated with little background. The prompt and displaced axion decays $a\to\gamma\gamma$ is another promising signature to probe an allowed region at $m_a<3m_\pi$. Since this channel was not explored in the previous work~\cite{Chakraborty:2021wda}, we study its potential here.

\section{Summary of $B \to K^{(*)}a$ theory calculation} 
\begin{figure}[t]
\includegraphics[width=0.25\textwidth]{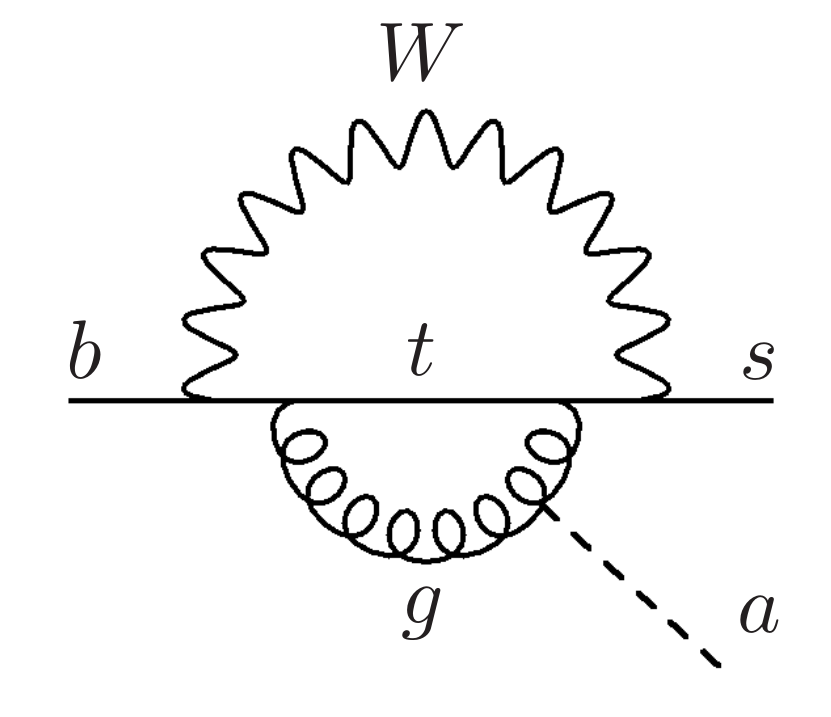}
\caption{A representative diagram of the leading contribution to the $b\to s a$ process from the Lagrangian~(\ref{eq:treelevellagrangian}).}  
\label{fig:2-loop-diagrams}
\end{figure}
Here we summarize the theoretical results of Ref.~\cite{Chakraborty:2021wda}. From the Lagrangian~(\ref{eq:treelevellagrangian}), the leading contribution to the process $b\to s a$ arises at 2-loop order, a representative diagram being shown in Fig.~\ref{fig:2-loop-diagrams}. The $b \to sa$ amplitude is captured by the following effective operator at scales below $M_W$: 
\begin{align}
\label{eq:matching}
\mathcal{L}_{bsa} = C \frac{\partial_\mu a}{f_a} \bar s_\text{\tiny L} \gamma^\mu \gamma_5 b_\text{\tiny L} + \text{h.c.}
\end{align}
with
\begin{align}
    \label{eq:Wilson}
    C = C_{bs}(\mu) + \frac{\alpha_w}{4\pi} C_{qq}(\mu) \, g(\mu)+ \frac{1}{2} \frac{\alpha_w}{4\pi}\left(\frac{\alpha_s}{4\pi}\right)^{\!2}\! f(\mu)\,.
\end{align}
Here, $\mu \sim M_W$, and we refer the reader to Appendix B of Ref.~\cite{Chakraborty:2021wda} for the (lengthy) expressions of the functions $f(\mu)$ and $g(\mu)$. 
The coefficients $C_{qq}(\mu)$ and $C_{bs}(\mu)$ are required by renormalization of the 2-loop diagrams, the former being the coefficient of the $(\partial_\mu a / f_a) \, \bar{q}\gamma^\mu\gamma_5 q$ counterterm, and the latter being that of $(\partial_\mu a / f_a) \, \bar{s}_\text{\tiny L} \gamma^\mu \gamma_5 b_\text{\tiny L}$.
Physically, these two coefficients parametrize the inevitably model-dependent effects of the UV physics that supersedes the low energy description, Eq.~(\ref{eq:treelevellagrangian}), above some high scale $\Lambda_\text{UV}$.
Rather than committing to a particular UV model, 
Ref.~\cite{Chakraborty:2021wda} varies the ``initial conditions'', $C_{qq}(\Lambda_\text{UV})$ and $C_{bs}(\Lambda_\text{UV})$, over their natural ranges and uses the renormalization group evolution of $C_{qq}(\mu)$ and $C_{bs}(\mu)$ to study the impact of the unknown UV physics on the bounds on $m_a$ and $f_a$ extracted in the infrared\@.
To go from the $b \to sa$ amplitude to the $B\to K^{(*)}a$ branching fraction, the result above is then combined with form factors obtained from the light-cone QCD sum rules~\cite{Ball:2004rg,Ball:2004ye,Izaguirre:2016dfi,Batell:2009jf}. 
An approximate formula for the branching fraction based on Eq.~(8) of Ref.~\cite{Chakraborty:2021wda} with $A=+3$ and $B=-3$ is given by
\begin{align}
    \label{eq:branching}
    \text{BR}\left(B^+ \to K^+ a\right) \simeq 2(10) \times 10^{-5} \left(\frac{100~\text{GeV}}{f_a}\right)^{\!\!2}\;,
\end{align}
for $\Lambda_{\text{UV}}=1(10)$~TeV, respectively. 

\section{Axion Decays: Displaced and Prompt}
%
\begin{figure}[t!]
\includegraphics[width=0.45\textwidth]{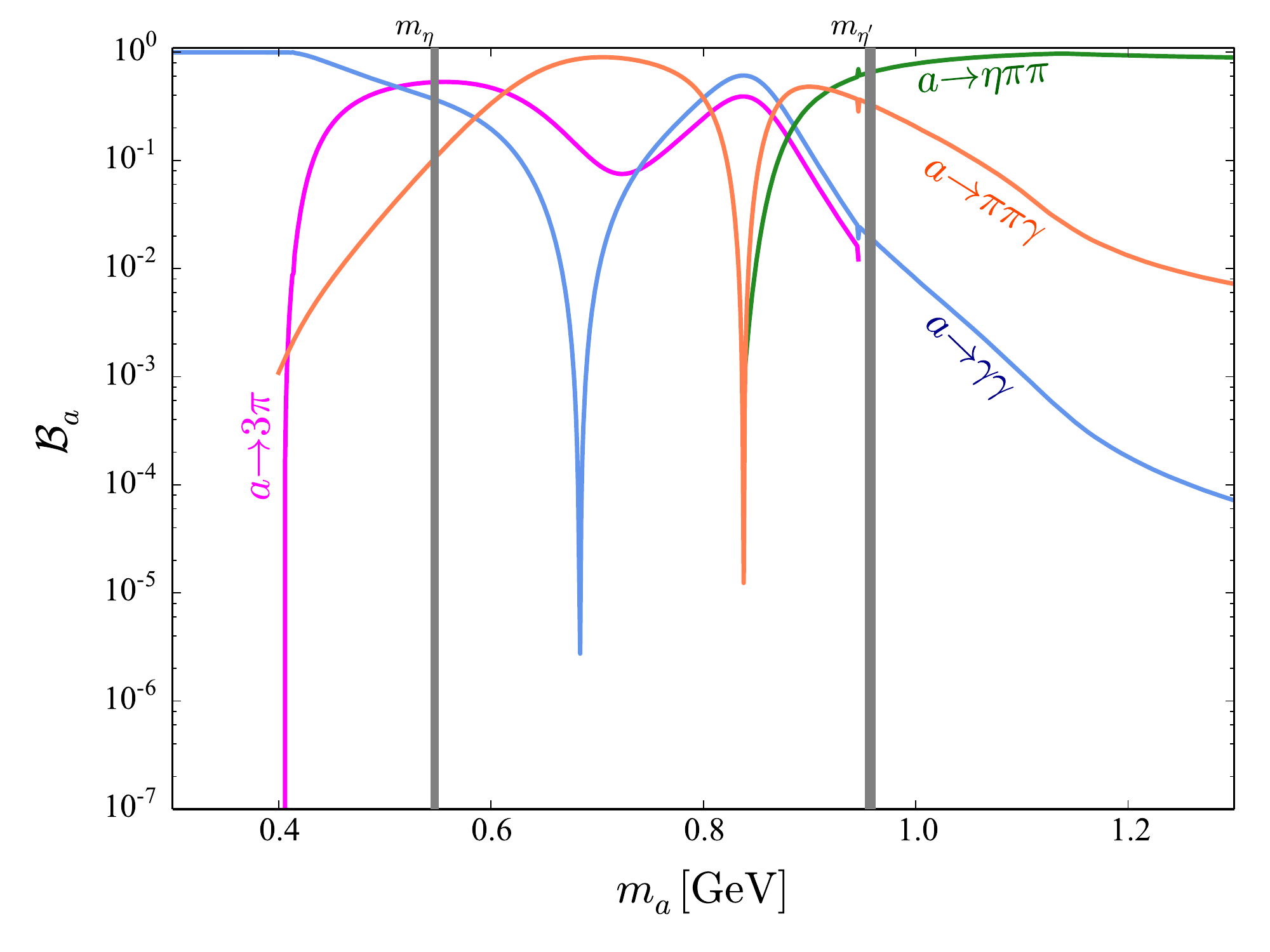}
\caption{Heavy QCD axion branching fractions as a function of its mass, taken from~\cite{Aloni:2018vki}. The decay modes relevant to our analyses are $a\to 3\pi$, $\gamma\gamma$, and $\eta\pi\pi$.}
\label{fig:branching}
\end{figure}

Due to its coupling to gluons, the decays of a heavy QCD axion are very diverse \cite{Aloni:2018vki}.
In terms of the ranges of $m_a$, we can summarize the decay patterns as follows.

For $3m_\pi < m_a < m_\eta+2m_\pi\sim 900$~MeV,
the branching fraction of $a\rightarrow 3\pi$ is sizable, as shown in Fig.~\ref{fig:branching}. 
In this mass range, with $f_a$ in the range of our interest $1\lesssim f_a \lesssim 10$~TeV, the axion can give rise to a displaced vertex signature in the Belle~II detector. Because of the displacement, one can optimize experimental cuts to reduce the background considerably. Consequently, analyzing displaced $B\to K a(3\pi)$ can result in very strong bounds on $f_a$. 
For higher $f_a$ the axion would be effectively invisible, a signature for which detection methods exist, but with very low efficiency~\cite{BaBar:2013npw}. Furthermore, as seen from Eq.~\eqref{eq:branching}, the production rate of the axion in this case is too small even with the full data set of Belle~II. 
Hence, the channel $B\to K a({\rm invisible})$ is unfavorable, unless other couplings than in Eq.~\eqref{eq:treelevellagrangian} increase the  axion production.

For $m_a>m_\eta+2m_\pi$, the decay mode of $a\to\eta\pi\pi$ quickly dominates, and  the axion lifetime becomes shorter. This channel was studied in Ref.~\cite{Chakraborty:2021wda}.

For $m_a < 3m_{\pi}$,  $a\rightarrow2\gamma$ dominates.%
\footnote{$a\rightarrow \gamma \pi\pi$ is allowed, but its branching ratio is very small for $m_a<3m_\pi$. See the orange line of Fig.~\ref{fig:branching}.}
Thus, to search for a heavy QCD axion with mass below $\sim400$ MeV one has to study the diphoton final state. 

Following Ref.~\cite{Aloni:2018vki}, we exclude few-MeV-wide regions of $m_a$ around $m_\eta$ and $m_{\eta'}$, because a large mixing of $a$ with $\eta$ or $\eta'$ would invalidate the perturbative treatment. 

To summarize, in this paper, we present projections for the reach of Belle~II using displaced $B\rightarrow Ka(3\pi)$ decays, and $B\rightarrow Ka(\gamma\gamma)$ decays that may be prompt or displaced.

\subsection{Displaced $B^-\to K^-a(\pi^+\pi^-\pi^0)$ signature}
\label{sectionA}

In the mass range $3m_\pi < m_a<m_\eta+2m_\pi$, we propose to search for the long-lived axion in $B^-\to K^-a$ with $a\to \pi^+\pi^-\pi^0$. 
The two charged pions form a displaced vertex (DV) significantly away from the interaction point of the $e^+e^-$ beams.
The DV is also the production point of the two photons that originate from the $\pi^0$ decay.
Following the usual practice of $B$-meson reconstruction at $e^+e^-$ $B$-factories, signal identification will rely on the variables $\Delta E\equiv E_B-\sqrt{s}/2$ and $M_{bc}\equiv \sqrt{s/4-p_B^2}$, where $s$ is the center-of-mass energy of the $e^+e^-$ collision, and $E_B$ and $p_B$ are the measured energy and momentum of the $B$ candidate in the center-of-mass frame.

\subsubsection{Background}

To estimate the background for this search, we start from the Belle Collaboration study of $B\to K\omega$ with $\omega\to \pi^+\pi^-\pi^0$~\cite{Chobanova:2013ddr}. Fig.~3b of Ref.~\cite{Chobanova:2013ddr} shows about 20 combinatorial-background events under the $B^-\to K^-\pi^+\pi^-\pi^0$ signal peak for $\pi^+\pi^-\pi^0$ mass in a 100-MeV-wide region and an integrated luminosity of about $0.75~\mathrm{ab^{-1}}$.
This corresponds to about 53,000 events for the Belle~II integrated luminosity of $50~\mathrm{ab^{-1}}$ and our $3\pi$ mass range, which is roughly $450<m_{3\pi}<950$~MeV. 
This is the expected background yield in a prompt search. 
However, exploiting the DV signature suppresses the background by more than a factor of $10^5$~\cite{Lees:2015rxq,Liventsev:2013zz,Lee:2018pag,Alimena:2019zri}. Therefore, we conclude that the combinatorial background in this search is well below 1 event throughout the entire $m_{3\pi}$ range.

In addition, a potential source of peaking background is $B\to K\omega$ and $B\to K\eta$ decays, with the $\omega$ and $\eta$ decaying to three pions.
Based on Refs.~\cite{Chobanova:2013ddr} and~\cite{Hoi:2011gv}, the numbers of events from these decays in a prompt analysis are similar to those of the combinatorial background.
From this, we conclude that after the DV requirement, these backgrounds become negligible as well. 
Therefore, there is no need to reject events with $m_{3\pi}$ around the $\omega$ or $\eta$ mass.

A more serious source of peaking background is $B^-\rightarrow K^-K_L$ with the long-lived $K_L$ decaying to $\pi^+\pi^-\pi^0$.
The branching fraction of this decay chain is about $8.1\times 10^{-8}$~\cite{ParticleDataGroup:2020ssz}.
Given the $K_L$ lifetime ($c\tau\approx 15$~m), the reconstruction efficiency is about 0.12\%.
This results in about 5 events in the Belle~II dataset. 
We use this estimated background to calculate a reduced sensitivity for  $m_a$ values in the 25-MeV-wide bin centered at 500~MeV. 
This bin is much wider than the  $m_{3\pi}$ resolution, which is only a few MeV~\cite{Belle:2007fdf}.

Peaking background may also arise from  $B^-\rightarrow K^-K^{*0}(892)$ with  $K^{*0}(892)\to K_S \pi^0$, and the long-lived $K_S$ decaying to $\pi^+\pi^-$ and forming a DV. 
The branching fraction for this decay chain is about $7\times 10^{-7}$~\cite{ParticleDataGroup:2020ssz}. 
This background is efficiently removed by rejecting events with $m_{\pi^+\pi^-}\approx M_{K_S}$~\cite{Lees:2015rxq} and for which the $\pi^+\pi^-$ momentum vector points from the interaction point to the DV.
The impact of this $K_S$ veto on the signal efficiency is small given the $m_{\pi^+\pi^-}$ resolution of about 4~MeV~\cite{Belle:2007goc}, the momentum angular resolution of order a milliradian~\cite{BaBar:2014omp}, and the DV position resolution of tens to hundreds of microns~\cite{BaBar:2014omp}, depending on the DV position. 
If needed, further suppression may be obtained by rejecting events for which the invariant mass of the $\pi^0$ with the displaced $\pi^+\pi^-$ pair, evaluated at the interaction point, is close to the peak of the $K^{*0}(892)$. 
Due to the $K^{*0}(892)$ width of about 50~MeV, this last cut would be more effective for low axion masses within our range of interest.
We note also that at the limit of the experimental sensitivity, corresponding to highly long-lived axions, this background is  exponentially suppressed by the relatively short ($c\tau\approx 2.7$~cm) lifetime of the $K_S$.
In our estimates we do not apply this requirement.

Thus, we conclude that the expected number of background events is below 1 event except for the 5 events in the region $m_{3\pi}\sim M_{K_L}$.
The background yield in any narrow range corresponding to a signal peak with width of order a few MeV is even lower.

\subsubsection{Efficiency}

\begin{figure}[h!]
\includegraphics[width=9cm,height=7cm]{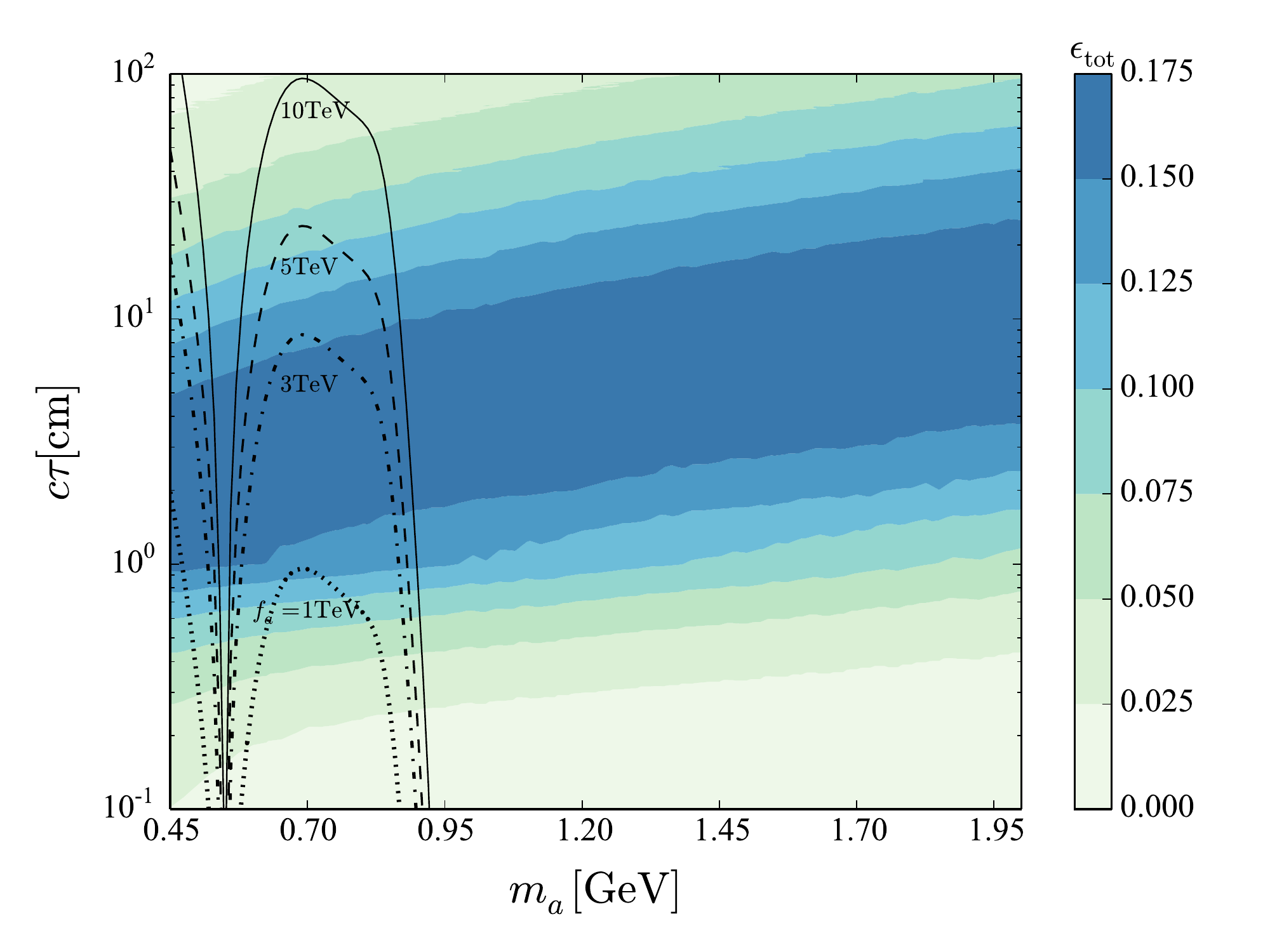}
\caption{Contours of the total efficiency for $B^-\to K^-a$ in the displaced $a\to\pi^+\pi^-\pi^0$ channel as a function of the axion mass and $c\tau$ where $\tau$ is the proper lifetime of the axion. 
}
\label{fig:efficiency}
\end{figure}

\begin{figure*}[!htbp]
\centering
\includegraphics[width=0.49\textwidth]{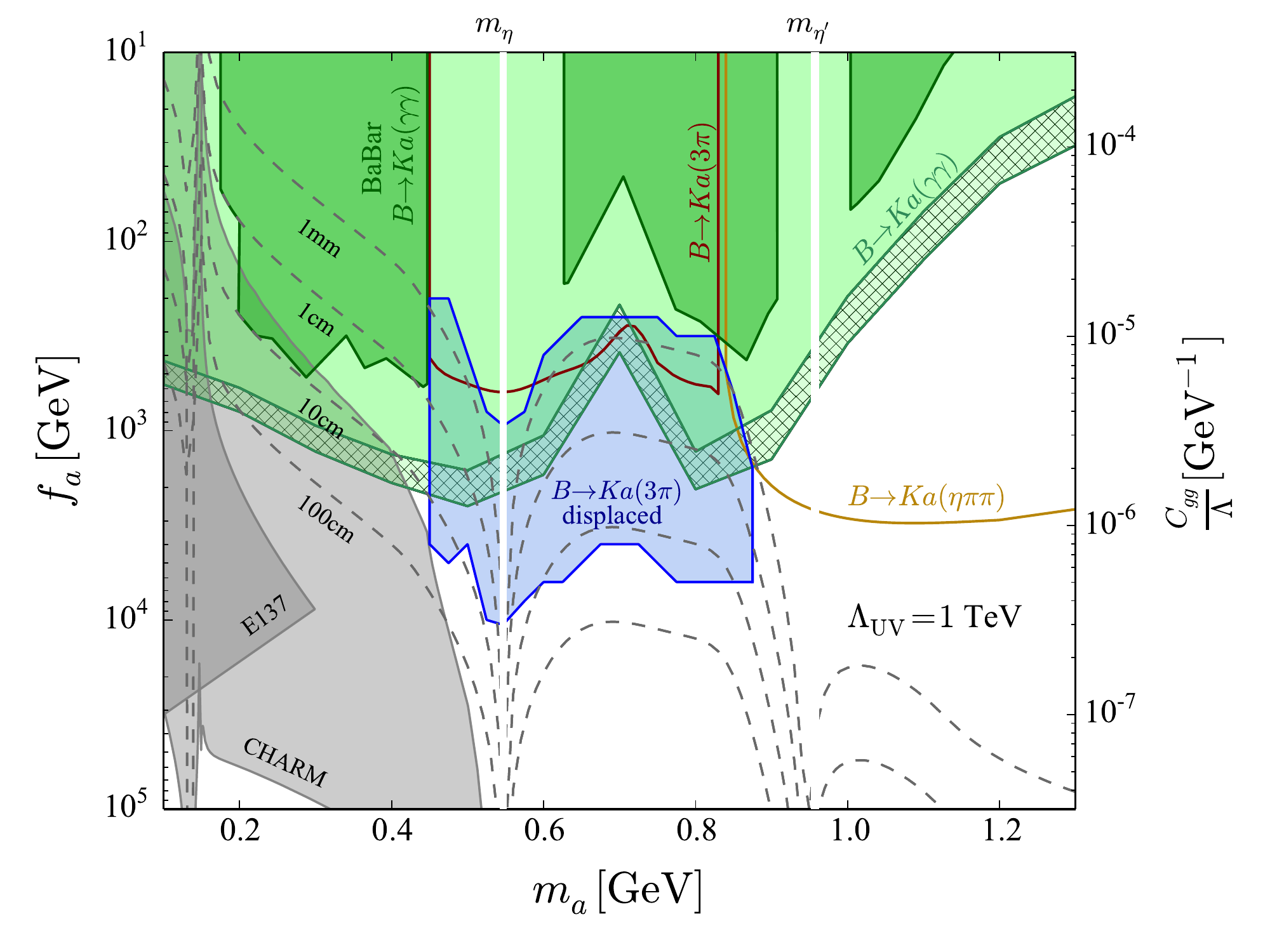}
\includegraphics[width=0.49\textwidth]{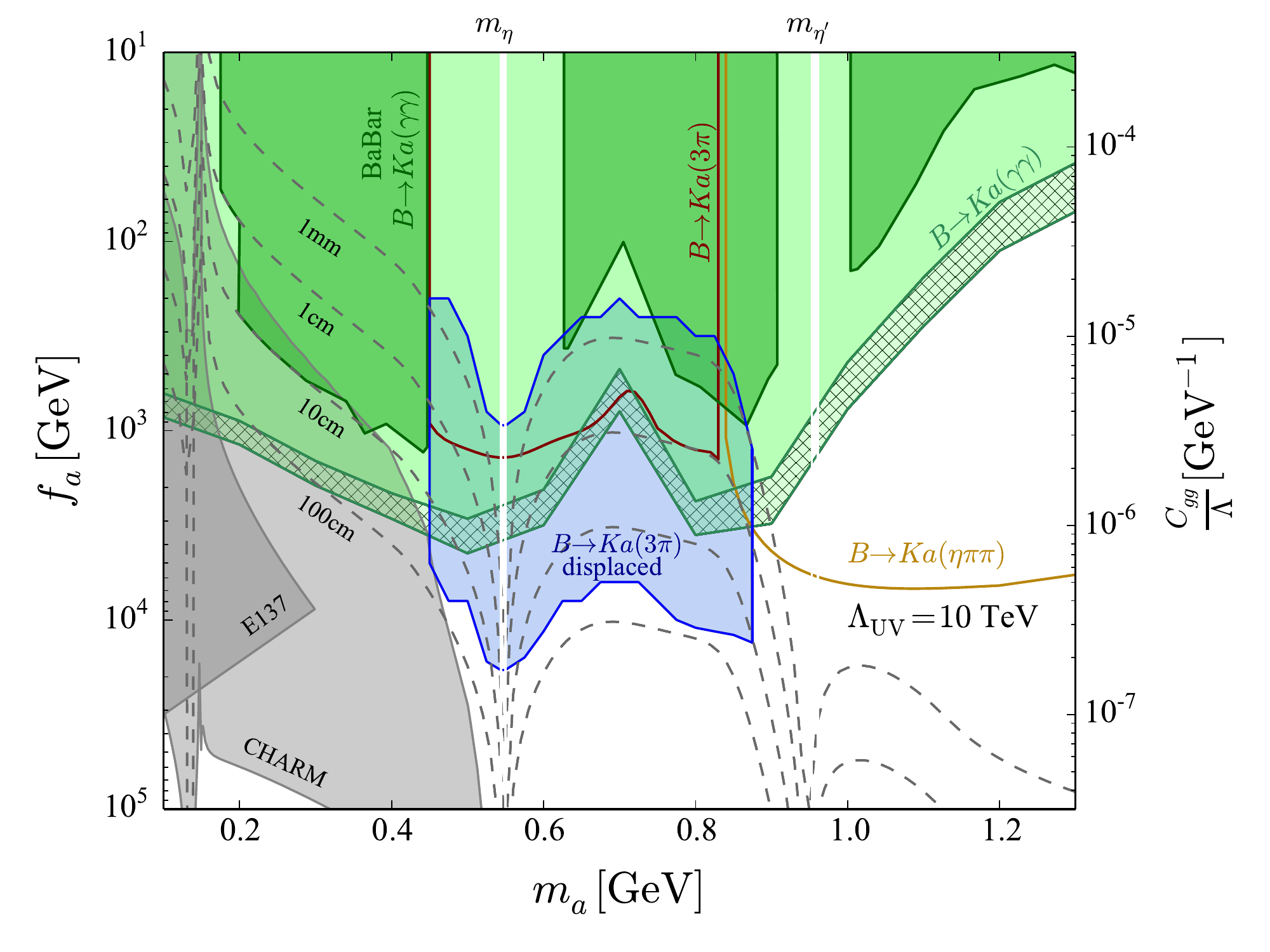}
\caption{In colors, we show the projected sensitivities and the new bound for $\Lambda_{\text{UV}}=1$ TeV (left figure) and 10 TeV (right figure) respectively, using prompt and displaced analysis. The projections developed in this paper are the  $B\to Ka(3\pi)$ {\it displaced} analysis (blue region) and $B\to Ka(\gamma\gamma)$ prompt analysis (green region). The green hatched region corresponds to a 10-fold variation of the estimated background rate in the $B\to Ka(\gamma\gamma)$ analysis. The bound of this channel is obtained by recasting the BABAR result~\cite{BaBar:2021ich} (dark green). 
The projections of $B\to Ka(3\pi)$ prompt analysis (magenta-outlined region) and $B\to Ka(3\pi)$ prompt analysis (yellow-outlined region) are from Ref.~\cite{Chakraborty:2021wda}. 
The grey regions refer to the present limits from $B$-decays, light meson decays and beam dump experiments.
The dashed contours show the axion's $c\tau$ values.}
\label{fig:projection}
\end{figure*}

To estimate the signal efficiency, we use EvtGen \cite{Lange:2001uf} to generate $B^-\to K^-a(\pi^+\pi^-\pi^0)$ events at the Belle~II beam energies ($E_{e^-} = 7$~GeV and $E_{e^+} = 4$~GeV). 
Simulated samples are produced for axion masses in range $450\le m_a \le 1950$~MeV in steps of 25~MeV. 
For Fig.~\ref{fig:efficiency}, samples are produced for 100 values of $c\tau \in [1{\rm mm}, 1{\rm m}]$. 
For calculation of the projected bounds, shown in Fig.~\ref{fig:projection}, samples are produced for 73 values of $f_a \in [1, 10^5]$~GeV, with $c\tau$ determined from $f_a$ according to Ref.~\cite{Aloni:2018vki}. 
Each sample contains $10^4$ events.
We calculate the effective efficiency for each sample as follows.
Following Refs.~\cite{Dib:2019tuj,Dey:2020juy}, we define the detector fiducial volume to be a cylinder of length  $-40 < z < 120$~cm along the beam direction and maximal radius $r < 80$~cm in the transverse plane, excluding the radial region $r<1$~cm in order to reject the promptly produced tracks. If a generated axion decays outside of the fiducial volume, its contribution to the efficiency is 0. For decays inside the fiducial volume, the radius-dependent track-detection efficiency $\epsilon_{\rm det}$ is taken to be linearly decreasing from $r=1$~cm ($\epsilon_{\rm det}=100\%$) to $r=80$~cm ($\epsilon_{\rm det}=0$)~\cite{Dib:2019tuj,Dey:2020juy}.
Finally, to take into account overall detection and reconstruction efficiencies, we multiply the efficiency by an overall factor of 22\%, which we estimate from the Belle study of $B \to K \omega$~\cite{Chobanova:2013ddr}, to obtain the total efficiency, $\epsilon_{\text{tot}}$. The total signal efficiency as a function of mean decay length and the axion mass is plotted in Fig.~\ref{fig:efficiency}. 

\smallsubsubsection{Projected bounds}
We estimate the Belle~II sensitivity to $B\rightarrow Ka( 3\pi)$ in terms of the 95\% confidence-level exclusion region in the plane of $f_a$ vs $m_a$. 
We assume that $5\times 10^{10}$ pairs of $B\bar B$ are produced given the integrated luminosity of approximately 50~ab$^{-1}$. 
Given that the displaced search is essentially background-free, we take the exclusion region to be that for which the number of signal events satisfies $N_S \ge 3$. At $m_a\simeq m_{K_L}$, we require $N_S \ge 10$, because 5 background events are expected. 
The excluded region is shown shaded blue in Fig.~\ref{fig:projection}, for the UV scales $\Lambda_{\text{UV}}=1$ TeV and $\Lambda_{\text{UV}}=10$ TeV.

\subsection{Prompt $B^-\to K^-a(\gamma\gamma)$ signature}
\label{sectionB}
In the mass range $m_a < 3m_\pi\simeq 400$~MeV, the axion decays predominantly to $\gamma\gamma$. 
In Sec.~\ref{sec:gg-recast}, we use a recent BABAR search for an axion-like particle (ALP) in this channel~\cite{BaBar:2021ich} to derive bounds on the parameter space of the heavy QCD axion.
In Sec.~\ref{sec:gg-projected}, we estimate the projected bounds at Belle~II, allowing for longer axion lifetimes than those reported in Ref.~\cite{BaBar:2021ich}. 

\smallsubsubsection{Recasting of BABAR results}
\label{sec:gg-recast}
Fig.~4 of Ref.~\cite{BaBar:2021ich} shows the bound on the branching-fraction product $\text{BR}(B\rightarrow Ka)\times \text{BR}(a\rightarrow\gamma\gamma)$ that BABAR obtained with an integrated luminosity of $424~{\rm fb}^{-1}$ as a function of $m_a$.
The bounds are shown for four values of axion lifetime, $c\tau_{\rm BBR}=0,0.1,1,10$~cm, and become weaker with increasing lifetime. 
To be conservative when recasting these  bounds, we take the axion with $m_a$, $f_a$, and corresponding $c\tau(f_a,m_a)$ values to be excluded if BABAR excludes the same values of $m_a$ and $\text{BR}(B\rightarrow Ka)\times \text{BR}(a\rightarrow\gamma\gamma)$ with an ALP lifetime that satisfies $c\tau_{\rm BBR}\geq c\tau(f_a,m_a)$. 
The resulting bounds, shown in dark green in Fig.\ref{fig:projection}, are naturally restricted by $c\tau<10$~cm.

\smallsubsubsection{Projected results for Belle~II}
\label{sec:gg-projected}

The $c\tau<10$~cm restriction, necessitated when recasting the results of Ref.~\cite{BaBar:2021ich}, is conservative given the size of the Belle~II calorimeter and the limited boost of the axion. 
Therefore, we also derive the expected Belle~II bounds from the $\gamma\gamma$ channel without this restriction, with the following procedure.

The impact of the background depends on the resolution of the signal peak in terms of the measured diphoton mass $m_{\gamma\gamma}$.  
For prompt axion decays, the detector  resolution ranges from $\sigma_{m_{\gamma\gamma}}\approx 11$~MeV for $m_{\gamma\gamma}\approx 540$~MeV~\cite{BelleIIRes} to $\sigma_{m_{\gamma\gamma}}\approx 50$~MeV for $m_{\gamma\gamma}\approx 1860$~MeV~\cite{BaBar:2011bxy}.
However, displaced photons have an additional source of smearing: the angular resolution of the calorimeter is not sufficient for determining the point of origin of a photon. Therefore, the diphoton mass $m_{\gamma\gamma}$ must be calculated assuming that the photons are promptly produced at the interaction point. 
This results in a downward smearing of the measured diphoton mass, 
$m_{\gamma\gamma}\simeq m_a(1-r/S)$, 
where $r=ct_a p_a/m_a$ is the flight distance of an axion with decay time $t_a$ and momentum $p_a$, and $S$ is the distance from the interaction point to the face of the calorimeter at the relevant point.
As a result, the exponential distribution of the flight distance translates into the $m_{\gamma\gamma}$ distribution 
\begin{align}
    \frac{dN}{dm_{\gamma\gamma}}=\frac{S}{p_a c\tau_a} \exp\left[\frac{S}{p_a c\tau_a}(m_{\gamma\gamma}-m_a)\right] \Theta(m_a-m_{\gamma\gamma}).
\end{align}
where $\Theta$ is the Heaviside step function. 
We take the typical values $p_a=2.5$~GeV and $S=120$~cm, and convolve this distribution with a Gaussian of width  $\sigma_{m_{\gamma\gamma}}\simeq 0.02 m_{\gamma\gamma}$, corresponding to the detector resolution~\cite{BelleIIRes, BaBar:2011bxy}. 
Our model reproduces very well the actual signal shape of the BABAR analysis for $m_a=1$~GeV and $c\tau_a=10$~cm, shown in p.18 of Ref.~\cite{shuveBabartalk}.

For given values of $m_a$ and $f_a$ we define the signal region to be the $m_{\gamma\gamma}$ region that contains $68\%$ of the signal around the peak. 
Following Ref.~\cite{BaBar:2021ich}, we take the efficiency for this decay chain to be 33\% and  calculate the expected signal yield $N_S$ in the signal region.  
The expected background yield in this region, $N_B$, is calculated given the width of the signal region and two values of the assumed background density: 150 events per MeV and 1500 events per MeV. 
These correspond to the background-level range shown in Fig.~1 of Ref.~\cite{BaBar:2021ich} for $m_a<1.3$~GeV. 
Since the sensitivity on $f_a$ scales as $N_B^{1/4}$, the 1-fold difference in background level has a relatively small impact on the projected limits. 
To avoid the background from $B\rightarrow K\pi^0(\gamma\gamma)$, $N_S$ and $N_B$ are calculated while excluding the mass range $125 < m_{\gamma\gamma} <145$~MeV, corresponding roughly to $\pm 2\sigma$ around the $\pi^0$ mass. 
Finally, we take the sensitivity of the experiment to be  $f_a$ and $m_a$ values for which $N_S/\sqrt{N_B}>2$.
The resulting projected limits are shown in light green in Fig.~\ref{fig:projection}.

\section{Discussion and summary}
In this paper we extract limits on the decay constant $f_a$ of the heavy QCD axion as a function of its mass $m_a$. 
We recast $B\to Ka(\gamma\gamma)$ results from BABAR~\cite{BaBar:2021ich}, and estimate the sensitivity of Belle~II for this decay using the BABAR efficiency and background and accounting for mass smearing due to displaced axion decays. 
as well as for a displaced, $B\to Ka( 3\pi)$ search. 

The projected sensitivities, shown in Fig.~\ref{fig:projection} are calculated for two UV scales, $\Lambda_{\text{UV}}=1$ TeV and $\Lambda_{\text{UV}}=10$ TeV. 
The sensitivity is higher for higher choices of the UV scale because of large logarithmic corrections originating from the renormalization group evolution.
We find that the dependence on the exact nature of the UV model, parametrized by the coefficients $A$ and $B$ in Ref.~\cite{Chakraborty:2021wda}, does not impact the results strongly.
The variation of $A$ and $B$ have sizeable effects only for $\Lambda_{\text{UV}}=1$ TeV, leading to at most  $O(1)$ variation in the limits on $f_a$. 
To avoid clutter in Fig.~\ref{fig:projection}, we chose  optimistic values, $A=+3$ and $B=-3$.

We find that for axion mass in the range $450 \lesssim m_a \lesssim 900$~MeV, the decay $a\rightarrow 3\pi$ with a displaced-vertex signature is the best search channel, with sensitivities to the axion decay constant in the range $10^2\lesssim f_a \lesssim 10^4$~GeV . 
Moreover, the  $a\rightarrow \gamma \gamma$ channel can be used to probe the mass range $150 \lesssim m_a\lesssim 500$~MeV, covering the unconstrained range $10\lesssim  f_a \lesssim 10^3$~GeV  of decay constant values.

\acknowledgements
We thank Brian Shuve for corresponding about Ref.~\cite{BaBar:2021ich}. SC, VL, TO, and KT are supported by the US Department of Energy grant DE-SC0010102. TO and KT are also supported in part by  JSPS KAKENHI 21H01086. EB and AS are supported by grants from the Israel Science Foundation, the US-Israel Binational Science Fund, the Israel Ministry of Science, and the Tel Aviv University Center for AI and Data Science.

%
\bibliographystyle{utphys}
\bibliography{references}
\end{document}